\documentclass[a4paper,11pt]{article}
\usepackage{pos}

\usepackage{graphicx}
\usepackage{subcaption}
\usepackage{caption}

\title{A Catalogue of Variable Active Galactic Nuclei Based on Multi-Timescale Variability Analysis from Fermi-LAT Data}
\ShortTitle{A Catalogue of AGNs Based on Multi-Timescale Variability from Fermi-LAT Data}

\author*[a]{Luana Passos-Reis}
\author[a]{Elisabete M. de Gouveia Dal Pino}
\author[b]{Tarek Hassan}
\author[c]{Santiago Pita}
\author[d]{Alberto Domínguez}

\affiliation[a]{Instituto de Astronomia, Geof\'{i}sica e Ci\^{e}ncias Atmosf\'{e}ricas (IAG-USP), Universidade de S\~{a}o Paulo, \\
Rua do Mat\~{a}o 1226, CEP: 05508-090, S\~{a}o Paulo - SP, Brazil.}

\affiliation[b]{Centro de Investigaciones Energéticas, Medioambientales y Tecnológicas (CIEMAT), 40, Madrid, Spain.}

\affiliation[c]{Université Paris Cité, CNRS, Astroparticule et Cosmologie, F-75013 Paris, France.}

\affiliation[d]{IPARCOS and Department of EMFTEL, Universidad Complutense de Madrid, E-28040 Madrid, Spain.}

\emailAdd{$^{*}$luana.passos.reis@usp.br}

\abstract{
Active Galactic Nuclei (AGN) sources feature supermassive black holes that launch relativistic plasma jets. They are key $\gamma$-ray sources that provide a unique laboratory for studying extreme particle acceleration and plasma physics. Variability in this $\gamma$-ray emission is an important signature that may constrain the size of the emission region and the underlying physical processes driving flares. However, current large-scale $\gamma$-ray catalogs, such as the 4LAC-DR3 catalogue \cite{Abdollahi_2020} from \textit{Fermi-LAT}, typically characterize variability only on long timescales (yearly or 60-day), lacking the necessary constraints on short-term behavior, from days to weeks. To address this gap, a systematic, high-cadence analysis is essential. We systematically characterize $\gamma$-ray variability in AGNs across short timescales: 3-day, 7-day (weekly), and 30-day (monthly). We present a preliminary catalogue of variable AGN based on $\gamma$-ray light curves from the Fermi-LAT Light Curve Repository \cite{Fermi_LCR}. Here we show that the variability amplitude ($\sigma_{\rm NXS}^{2}$) presents similar values across the different timescales, potentially increasing for a subsample of sources as the observation timescale increases. This high-cadence analysis reinforces the known dichotomy between flat-spectrum radio quasars (FSRQs) and BL Lacertae objects (BL Lacs), with FSRQs consistently exhibiting stronger variability.
By identifying the most luminous and variable sources at each timescale, we highlight key targets for observation strategies and follow-up observations with next-generation observatories such as the Cherenkov Telescope Array Observatory (CTAO), ASTRI Mini-Array, and the Southern Wide-field Gamma-ray Observatory (SWGO), where strong short-term variability suggests highly compact emission zones and extreme particle acceleration efficiency. This catalogue thus contributes to the physical understanding of high-energy outflows in AGN jets and provides a foundation for optimizing observational strategies through the development of a unified variability metric across timescales.
}

\FullConference{
High Energy Phenomena in Relativistic Outflows IX (HEPRO-IX)\\
4-8 August 2025\\
Rio de Janeiro, Brazil\\}

\begin{document}
\maketitle

\section{Introduction: Blazar Variability and CTAO}
\label{sec:intro}

Blazars are highly variable active galactic nuclei (AGN) with emission dominated by relativistic jets pointed at Earth. Their variability across timescales encodes important information about particle acceleration, cooling mechanisms, and the size of the emission region. In the Very High Energy (VHE) regime, short-timescale flux variations can significantly increase the chances of detection, during a flaring state, making short-timescale analysis essential for optimizing observations with next-generation instruments and observatories.

Existing public Fermi-LAT catalogs typically characterize variability on yearly or 60-days timescales \cite{Abdollahi_2020}. Our work addresses this gap by systematically characterizing $\gamma$-ray variability across shorter cadences: 30-day, 7-day, and 3-days, leveraging the Fermi-LAT Light Curve Repository \cite{Fermi_LCR}. This comprehensive short-term characterization directly supports the extragalactic science program of the Cherenkov Telescope Array Observatory (CTAO), where these results are used for population forecasts (\cite{PassosReis_2025} and CTAO Consortium, in prep.).

\section{Methodology: Data Sample and Normalized Excess Variance}

We analyzed 1429 AGN $\gamma $-ray light curves sourced from the Fermi-LAT Light Curve Repository \cite{Fermi_LCR}. Sources were sampled based on the availability of light curve data and analyzed using cadences of 3-day, 7-day (weekly), and 30-day (monthly). For contextual Spectral Energy Distribution (SED) classification, the full sample of sources from the 4LAC-DR3 Catalog \cite{Abdollahi_2020} lists 2825 blazars with defined peaks: 1699 Low (LSP), 536 Intermediate (ISP), and 590 High (HSP) blazars, reflecting the Fermi-LAT's energy range sensitivity bias towards LSPs.

\begin{figure}[ht]
    \centering
    \begin{subfigure}[t]{\textwidth}
        \centering
        \includegraphics[width=0.8\textwidth]{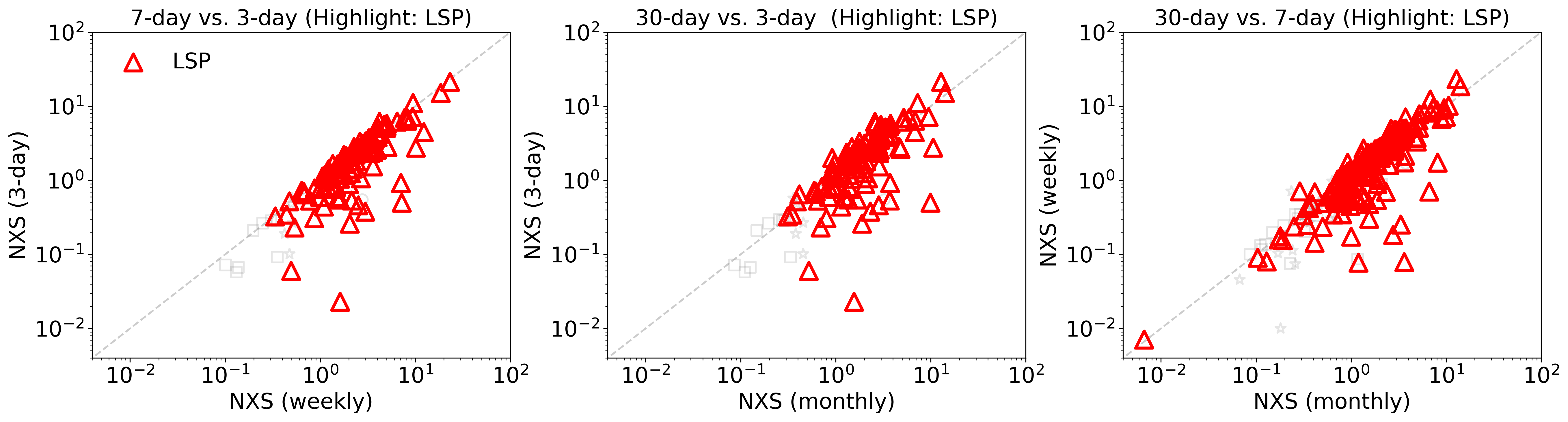}
    \end{subfigure}
    \begin{subfigure}[t]{\textwidth}
        \centering
        \includegraphics[width=0.8\textwidth]{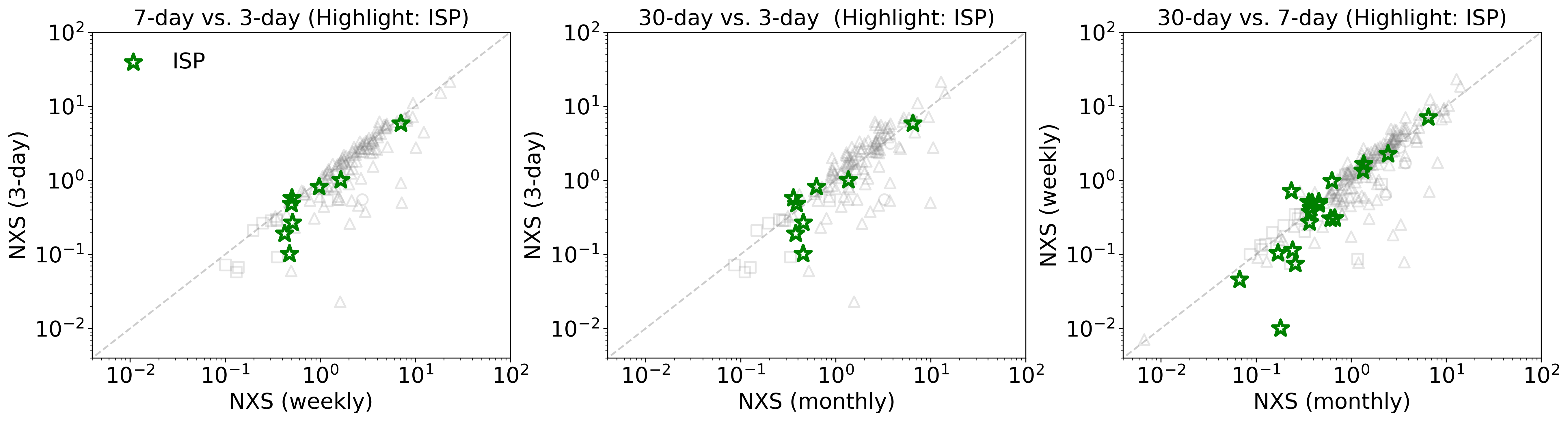}
    \end{subfigure}
    \begin{subfigure}[t]{\textwidth}
        \centering
        \includegraphics[width=0.8\textwidth]{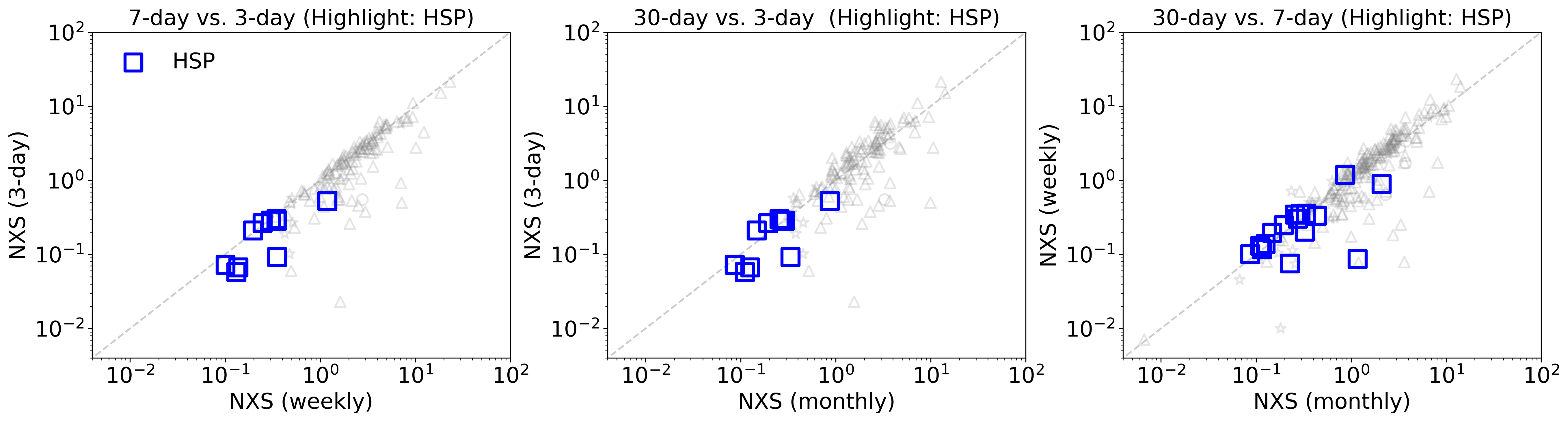}
    \end{subfigure}
    \caption{Distributions of the \textit{Normalized Excess Variance ($\sigma_{\rm NXS}^{2}$, NXS)} for sources in the catalogue, illustrating the correlation between variability amplitude across three distinct temporal regimes. The figure is structured as a $3 \times 3$ matrix of NXS correlations:
    \textbf{Rows (Synchrotron Peak Class):} The three rows highlight the source populations (Top to Bottom): \textit{Low Synchrotron Peak (LSP)} blazars (red triangles), \textit{Intermediate Synchrotron Peak (ISP)} blazars (green stars), and \textit{High Synchrotron Peak (HSP)} blazars (blue squares). Note the larger count of LSP sources, consistent with Fermi-LAT's sensitivity.
    \textbf{Columns (Timescale Comparison):} The columns (Left to Right) compare NXS estimates derived from the light curve binnings: 7-day vs. 3-day; 30-day vs. 3-day; and 30-day vs. 7-day. For all panels, the X-axis represents the longer cadence and the Y-axis represents the shorter cadence. The faint grey 1:1 line is included in all plots, representing the case where $\sigma_{\rm NXS, \text{shorter}}^{2} = \sigma_{\rm NXS, \text{longer}}^{2}$, nearly all data points present similar values across timescales, with a subset falling within a systematic distribution below the line, suggesting $\sigma_{\rm NXS}^{2}$ increasing with timescale ($\sigma_{\rm NXS, \text{shorter}}^{2} < \sigma_{\rm NXS, \text{longer}}^{2}$).}
    \label{fig:nxs_correlations}
\end{figure}


The analyzed data was quality-filtered using the outlier treatment detailed in \cite{PassosReis_2025} and by applying a Test Statistic ($\text{TS}$) threshold of $\text{TS} > 4$ (Passos Reis et al., in prep.). The amplitude of intrinsic variability is quantified using the \textbf{Normalized Excess Variance} (NXS, $\sigma_{\rm NXS}^{2}$) \cite{Vaughan_2003}, our primary metric in this work:

\begin{equation}\label{eq:nxs}
    \sigma_{\rm NXS}^{2} = \frac{1}{F_{\rm av}\ ^{2}} \left [\frac{1}{N - 1} \sum_{i=1}^{N} \left ( F_{i} - F_{\rm av} \right )^{2} - \frac{1}{N} \sum_{i=1}^{N} \sigma_{\rm err, i}^{2} \right ],
\end{equation}



\noindent where $N$ is the number of valid flux measurements, $F_{i}$ the flux in bin $i$, $F_{\rm av}$ the average flux, and $\overline{\sigma_{\rm err}^{2}}$ the mean squared flux error. The fractional variability amplitude, $F_{\rm var} = \max{ \left( 0, \sqrt{\sigma_{\rm NXS}^{2}} \right) }$, is used to ensure statistical robustness.

\begin{table}[h]
    \centering
    \caption{Number of sources from the 1429 AGN sample \cite{Abdollahi_2020,Fermi_LCR, PassosReis_2025} selected for plotting in Figure 1. Sources are included only if they passed the outlier treatment described in \cite{PassosReis_2025}, and yielded a statistically significant positive Normalized Excess Variance ($\sigma_{\rm NXS}^{2} > 0$) for \textit{both} cadences in a given comparison. This selection highlights the most reliably variable sources in the catalogue, and most promising targets for upcoming facilities.}
    \label{tab:plotted_source_counts}
    \begin{tabular}{|l|c|c|c|}
        \hline
        \textbf{SED Class} & \multicolumn{3}{c|}{\textbf{Number of Plotted Sources}} \\
        \cline{2-4}
        & \textbf{7-day vs. 3-day} & \textbf{30-day vs. 3-day} & \textbf{30-day vs. 7-day} \\
        \hline
        Low Synchrotron Peak (LSP) & 113 & 114 & 181 \\
        Intermediate Synchrotron Peak (ISP) & 8 & 8 & 20 \\
        High Synchrotron Peak (HSP) & 11 & 11 & 16 \\
        \hline
    \end{tabular}
\end{table}

The $\sigma_{\rm NXS, 3day}^{2}$ estimate was also expanded using a proportionality coefficient derived from the 30-day timescale in our previous work \cite{PassosReis_2025}, successfully increasing the short-timescale sample from 87 to 407 sources, used as a proxy for modeling daily variability in the CTAO AGN Population Task Force studies (CTAO Consortium, in prep.).

\section{Results and Discussion: Timescale Correlation}
\label{sec:results}

Our analysis confirms the known correlation between variability amplitude and timescale, 
reinforcing the dichotomy with flat-spectrum radio quasars (FSRQs) exhibiting consistently stronger variability than BL Lacertae objects (BL Lacs), supporting the characterization of high-energy outflows in AGN jets. As Figure \ref{fig:nxs_correlations} illustrates, the Normalized Excess Variance ($\sigma_{\rm NXS}^{2}$) values for the majority of sources are concentrated overlaid into the 1:1 line, visually suggesting similar short- and long-term variability.
However, the strict absence of sources above the 1:1 line, combined with a systematic tail composed by a subset of sources, extending below the grey line, demonstrates that $\sigma_{\rm NXS}^{2}$ increases with the observation timescale ($\sigma_{\rm NXS, \text{shorter}}^{2} \le \sigma_{\rm NXS, \text{longer}}^{2}$) for a given set.
The resulting slope and dispersion observed in these NXS-timescale correlations may reflect class-dependent variability mechanisms:


\begin{itemize}
    \item FSRQs (LSP dominance) exhibit the strongest and broadest variability, consistent with emission dominated by External Compton (EC) processes.
    
    \item BL Lacs span all classes, including all HSP blazars and most of the ISP blazars seen in our source sample \ref{tab:plotted_source_counts}, as well as some less variable LSPs,
    reflecting the dominance of Synchrotron Self-Compton (SSC) processes. Their strong,
    episodic flaring events highlights the role of instruments like CTAO in detecting high-energy transients. ISP sources may represent a transition zone between persistent and episodic variability.
\end{itemize}

This work identifies subsamples of the most luminous and variable sources (Table \ref{tab:plotted_source_counts}) as primary targets for next-generation VHE observatories (CTAO, ASTRI Mini-Array, SWGO). Strong variability reveals compact emission regions and extreme particle acceleration, optimizing transient-focused observing strategies. The resultant catalogue (Passos Reis et al, in prep.) serves as a foundational step toward developing a unified variability metric. Ongoing work will integrate variability-informed extrapolations into the CTAO Scientific Collaboration's forecasts (CTAO Consortium, in prep.), with future efforts including yearly timescale analysis and complementary Swift X-ray data.



\section*{Acknowledgements}
The authors acknowledge the important participation of Jean Philippe-Lenain and Jonathan Biteau, as well as the members of the CTAO AGN Population Task Force. This work is supported by ongoing efforts for the AGN Population CTAO Consortium paper (in prep.). This work has received partial support from FAPESP (grants n. 2021/02120-0 and 2024/05459-6).

\end{document}